\documentclass[aps,pre,twocolumn,groupedaddress,showpacs]{revtex4}
\usepackage{amssymb}
\usepackage{graphicx,color}
\definecolor{r}{rgb}{1,0,0}
\definecolor{g}{rgb}{0,1,0}
\definecolor{b}{rgb}{0,0,1}

\begin{document}


\title{Propagating Waves in a Monolayer of Gas-Fluidized Rods}


\author{L.J. Daniels and D.J. Durian}
\affiliation{Department of Physics and Astronomy, University of
Pennsylvania, Philadelphia, PA 19104-6396, USA}


\date{\today}

\begin{abstract}
We report on an observation of propagating compression waves in a quasi-two-dimensional monolayer of apolar granular rods fluidized by an upflow of air. The collective wave speed is an order of magnitude faster than the speed of the particles. This gives rise to anomalously large number fluctuations $\Delta N \sim  N^{0.72\pm0.04}$, which are greater than ordinary number fluctuations of $N^{1/2}$. We characterize the waves by calculating the spatiotemporal power spectrum of the density. The position of observed peaks, as a function of frequency $\omega$ and wavevector $k$, yields a linear dispersion relationship in the long-time, long-wavelength limit and a wavespeed $c = \omega/k$.  Repeating this analysis for systems at different densities and air speeds, we observe a linear increase in the wavespeed with increasing packing fraction with no dependence on the airflow. Although air-fluidized rods self-propel individually or in dilute collections, the parallel and perpendicular root-mean-square speeds of the rods indicate that they no longer self-propel when propagating waves are present.  Based on this mutual exclusivity, we map out the phase behavior for the existence of waves vs self-propulsion as a function of density and fluidizing airflow.
\end{abstract}



\maketitle



\section{Introduction}

Macroscopic granular systems have been studied extensively due to their ubiquity in industrial applications as well as the ease with which they may be visualized~\cite{BrownRichards,Nedderman,JNBrev96}. Furthermore, air-fluidized grains have demonstrated a remarkable ability to serve as both a system analogous to Brownian particles~\cite{Durian2004} as well as a proxy for colloidal and molecular glasses~\cite{Abate2006, AbatePRL08}. Idealized granular media consist of spherical particles whereas real systems will have both polydispersity and anisotropic particle shape. The role of particle shape is now a topic of increasing attention. Recent experiments on vibrated rods~\cite{Kudrolli2003, Olafsen2005, Nossal2006, Menon2006, Tsimring2007} and rods in a hopper~\cite{Franklin2009} have demonstrated the great variety of phenomena -- swirling, vortices, pattern formation -- that arise due to particle anisotropy alone.

Individual air- or vibro-fluidized granular rods have also been observed to self-propel, in the sense of moving preferentially along their long axis~\cite{Menon2007, TsimringPolar, Daniels2009}. The idea of active self-propulsion has garnered great theoretical interest \cite{TonerAP05, RamaswamyARCM10}  due to the potential of unifying a broad spectrum of systems over a huge range of length scales: from biological systems like bacteria~\cite{Bar2002, Kessler, Oster2004, Kaiser2001}, fish~\cite{Fish}, locusts~\cite{Locusts}, birds~\cite{Starlings, Birds}, to physical systems such as agitated granular materials~\cite{Kudrolli2003,Tsimring2007} and human or animal traffic~\cite{Traffic, Ants}. In this context, self-propelling at a microscopic scale gives rise to dramatic collective behavior at the macroscopic scale including the emergence of a dynamic broken-symmetry state in which all the particles move in the same spontaneously-chosen direction as a coherent flock~\cite{Vicsek1995, Tu1995, Tu1998}. These systems are also predicted to exhibit collective motion, propagating waves, and anomalously large number fluctuations~\cite{Ulm1998, Simha2002, Ramaswamy2003, Chate2006, Marchetti2008}. Despite the wealth of theory and simulation, very few experiments have been conducted due to both the difficulty of tracking and analyzing biological systems as well as finding purely physical systems in which particles self-propel. Of note, giant number fluctuations were observed in a vertically-vibrated monolayer of granular rods possessing nematic order~\cite{Menon2007}. Additionally, collective behavior in the form of whorls, jets, and vortices -- in which the collective motion was an order of magnitude faster than individual particle motion -- has been observed in dense populations of bacteria~\cite{Kessler}.

For this paper, we extend our earlier experiments on individual and dilute systems of air-fluidized rods~\cite{Daniels2009} in order to search for giant number fluctuations and collective flock-like motion.  In spite of considerable effort, we never found conditions under which flocking occurred.  We do observe giant number fluctuations, however, and we additionally find that they appear to be coupled to the unexpected existence of density waves that propagate ballistically across the system.  To begin, in Section~\ref{exp}, we introduce the apparatus and system under study. The paper will first examine a single system at fixed density of 64\% packing fraction. We describe our observation of propagating compression waves in Section~\ref{pwaves}. The wavespeed is an order of magnitude larger than the root-mean-square speed of the particles; this separation of timescales results in giant fluctuations in the local number density, discussed in Section~\ref{nflucts}. Lastly, in Section~\ref{quantwaves}, we quantify the waves by calculating the spatiotemporal power spectrum of the density and extracting a dispersion relationship for the waves. We extend the experiments by examining how the wavespeed varies with density and airflow and mapping out phase behavior.

\begin{figure*}
\includegraphics[width=7.00in]{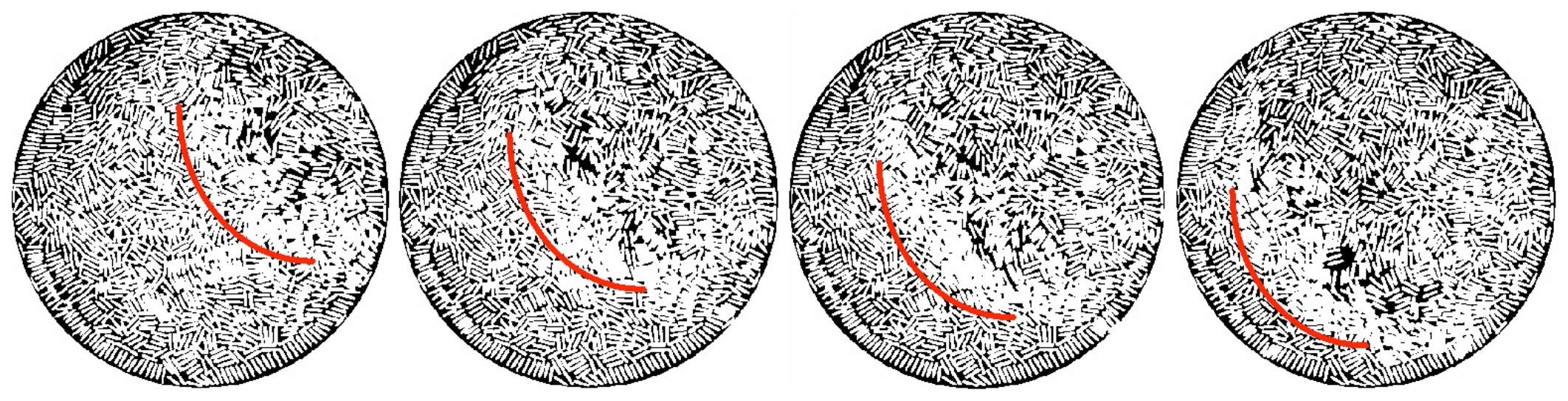}
\caption{(Color online) A propagating wave in a monolayer of gas-fluidized rods. The system has diameter 28.6 cm. The rods are 1.27~cm long, have an aspect ratio of five, occupy 64\% projected-area packing fraction, and are fluidized by an upflow of air at 220~cm/s. The time between images, moving left to right, is 0.15~s.  The red arc is moved at constant velocity 28 cm/s  as a guide to the eye.}
\label{WAVES}
\end{figure*}

\section{Experimental Details}\label{exp}

We study a monolayer of right cylindrical acetal dowel rods -- length 1.27~cm, diameter 0.24~cm, aspect ratio 5, and mass 0.076~g -- fluidized by an upflow of air. Except when otherwise noted, we analyze a system of 1353 rods occupying packing fraction 64\%, fluidized by a spatially and temporally uniform upflow of air at speed 220 cm/s as measured by a hot-wire anemometer. The fluidizing air speed is low enough that rods do not overlap out-of-plane and the system remains a monolayer.

The apparatus and fluidization method is identical to that of Ref.~\cite{Daniels2009}. The apparatus is a rectangular windbox, $0.5 \times 0.5 \times 1.2~{\rm m}^3$, positioned upright. A circular brass testing sieve with mesh size 150~$\mu$m and diameter 30.5~cm rests horizontally on top.  In order to prevent particles from becoming trapped in a small groove around the edge of the sieve, we place a 1~cm norprene tube around its inside edge. A blower attached to the windbox base provides vertical airflow perpendicular to the sieve. Raw video data of the fluidized particles is captured for 10 minutes at 120 frames per second by a digital camera mounted above the apparatus. Post-processing of the video data is accomplished in LabVIEW, using the same tracking programs used in Ref.~\cite{Daniels2009} when appropriate. For the analysis of the waves in Section~\ref{quantwaves}, we treat the video as a binary density map $\rho (x,y,t)$ rather than individually track particles.

In order to emphasize the role that shape plays here, we analyze a companion system of 400 bidisperse plastic spheres -- diameters 0.64 cm and 0.95 cm -- chosen to have the same packing fraction 64\% as the rod system. The fluidization speed, 260 cm/s, for the sphere system is chosen so that the the spheres and rods have roughly the same rms speed.  This is below the terminal falling speed, so the spheres roll without slipping or levitating.

Prior to focussing on the 1.27~cm rods with aspect ratio 5, we also explored a number of other granular systems.  This includes apolar rods with aspect ratios of 4, 9, 14, and 30; polar rods having a distinct head and tail with aspect ratio 2.5 and 14; a range of airspeeds from 10 to 500~cm/s; and densities from 10\% to 85\% area packing fraction.  We also examined such systems in a circular race track with width that varied from approximately three to ten particle widths, by placing a cylinder concentrically onto the sieve.  In no case did we observe flocking.

\section{Propagating Waves}\label{pwaves}

We observe a propagating compression wave instability for dense collections of fluidized rods. This phenomenon is unique to rods; we do not observe compression waves for spheres. A time series of images, each separated by 0.09 s, shown in Figure~\ref{WAVES} depicts one such propagating compression wave for a system at 67\% packing fraction. The red arc moves at constant velocity 28 cm/s with the wavefront and serves as a guide to the eye. When particles are compressed together, light is specularly reflected from the interface and the wavefront appears brighter in the images and the less dense rarefaction zones darker.  An example video clip is available on-line \cite{SOOM}.

\begin{figure}
\includegraphics[width=3.25in]{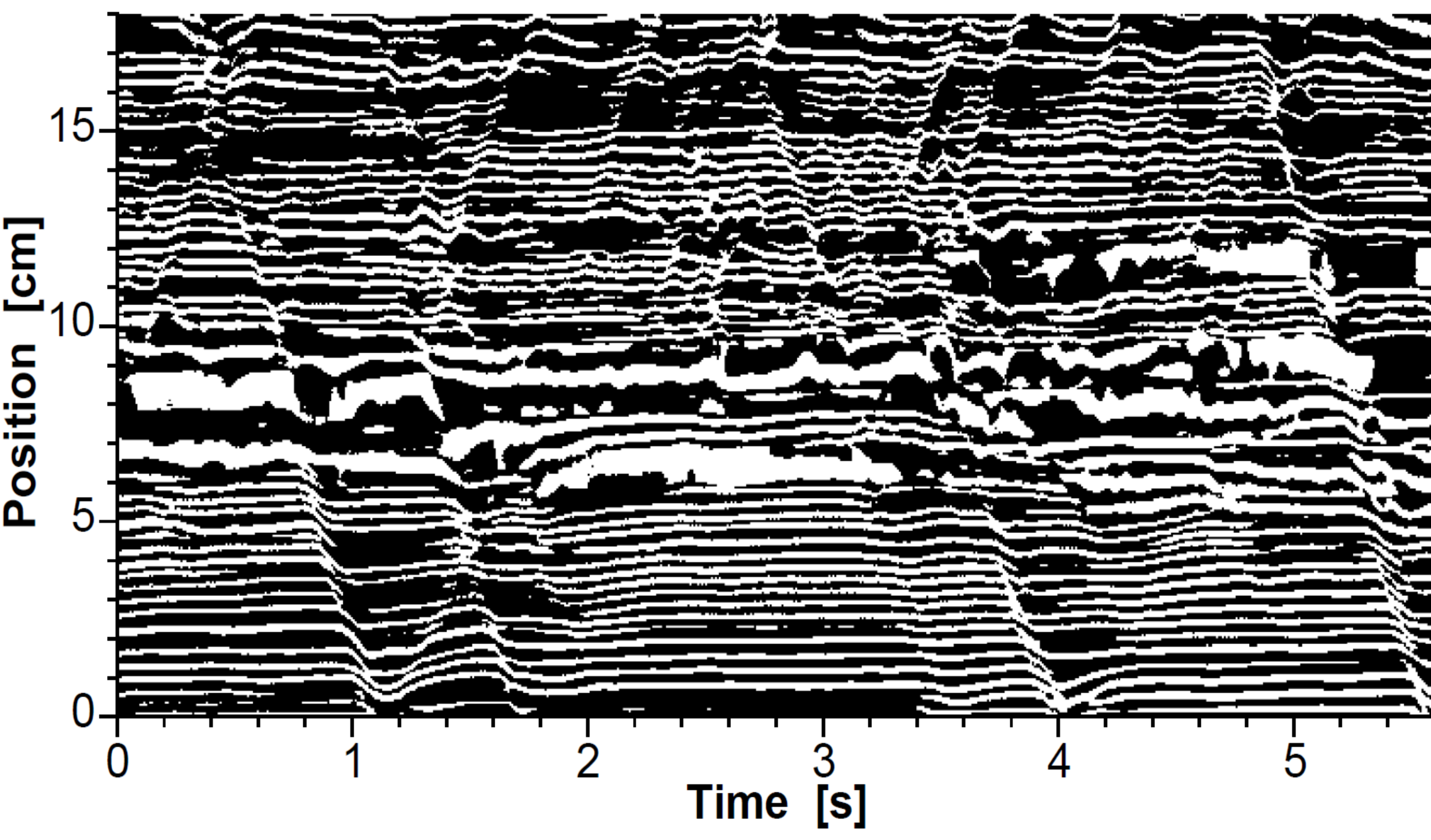}
\caption{Spacetime plot of a line of pixels taken from a video of air-fluidized rods as specified in Section~\ref{exp}.}\label{ST}
\end{figure}

As can be seen in videos, e.g.~\cite{SOOM}, the waves travel much faster than the particles themselves. The waves propagate through the medium regardless of the local ordering with some preference for moving along the local splay direction. To get an estimate for the wavespeed without any detailed processing, we look at spacetime plots of a single line of pixels extracted from the video data, as shown in Fig.~\ref{ST}. The waves are visible on the spacetime plot as diagonal lines. From the slopes of these lines, we obtain a range of wavespeeds from 10 to 30~cm/s, in agreement with the moving arc in Fig.~\ref{WAVES}. From position-versus-time data for the individual rods, we calculate the root-mean-square particle speed of $\sim$1~cm/s. Thus, the waves move an order of magnitude faster than the particles themselves and an order of magnitude slower than the fluidizing airflow. This is similar to what has been observed for collective motion of bacteria in which the collective behavior is an order of magnitude faster than the individual particle speeds~\cite{Kessler}. However, both a hydrodynamic theory of particles adsorbed on a substrate~\cite{Mazenko1982} as well as a microscopic theory of self-propelling rods~\cite{Marchetti2008} predict that propagating waves in those systems have a speed $c$ equal to $v_{rms}$.

\begin{figure}
\includegraphics[width=3.25in]{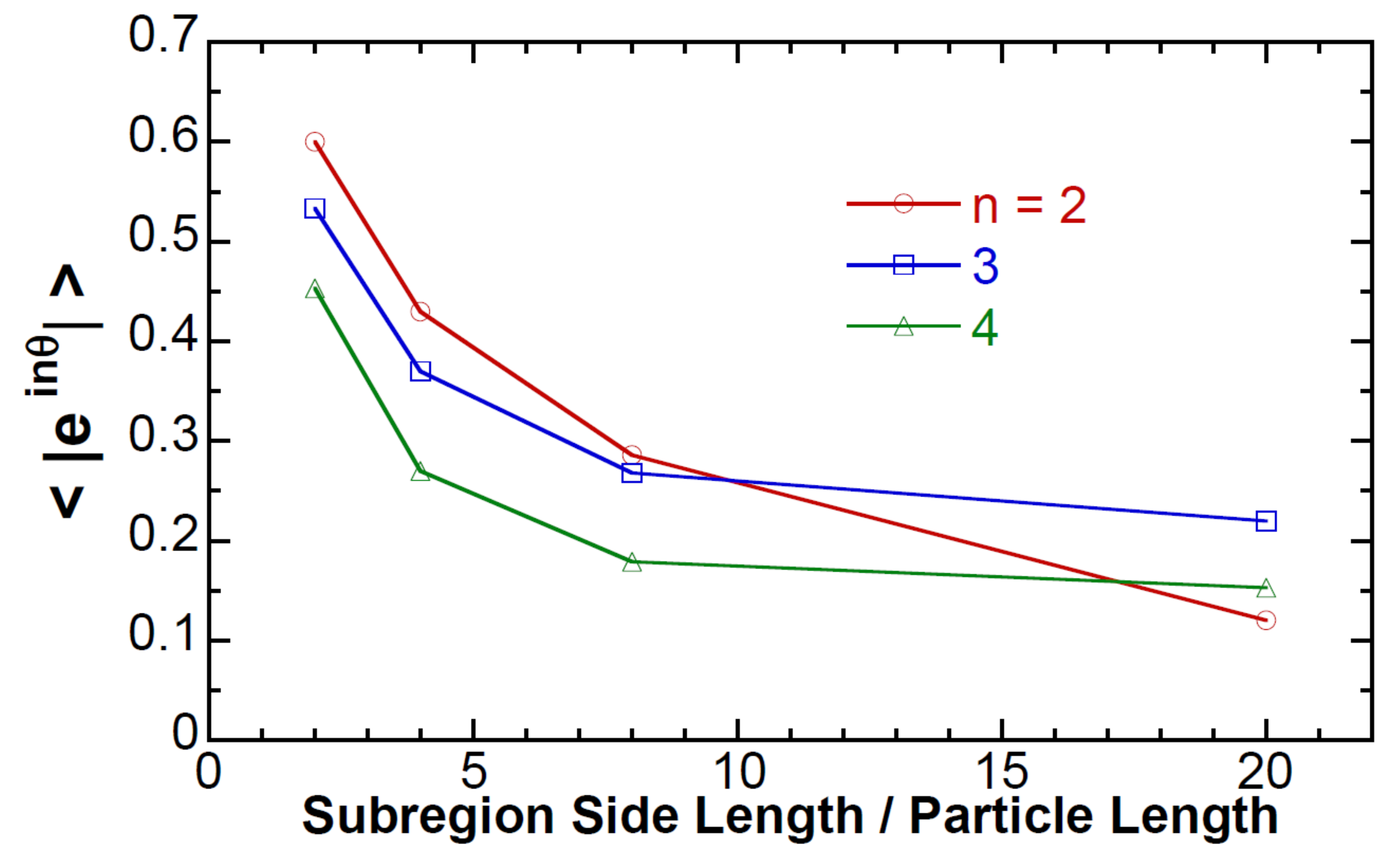}
\caption{(Color online) Time- and ensemble-averaged orientational order parameter $\langle |e^{in\theta}| \rangle$ for $n$= 2, 3, 4 for a system of air-fluidized rods at 64\% packing fraction. The ensemble average is taken over all particle orientations within a square subregion with side lengths $l \times l$. The order parameter is plotted against this side length scaled by the rod length.}\label{EINQ}
\end{figure}

Predictions of waves for self-propelling rods typically require that long-range order or a dynamic broken symmetry state be present in the system. By observation, neither appears to develop for our systems. The emergence is suppressed by the propagation of the compression waves which disrupt the local ordering and any coherent motion developing between neighboring particles. However, although there is no long-range order, localized domains of nematic ordering are observed and the particles in these regions tend to move in the same direction in a bulk sense -- we observe localized correlated motion. To quantify the extent of these local domains, we calculate the ensemble- and time-averaged orientational order parameter $\langle |e^{in\theta}|\rangle$ for $n =$ 2, 3, 4. For a fixed packing fraction and airflow, we vary the neighborhood size over the which the ensemble average of orientations is taken. The neighborhood is a square subregion with side lengths $l \times l$. The result for a system at 64\% packing fraction is shown in Fig.~\ref{EINQ} where the subregion side length has been scaled by the rod length. If a system possesses an $n$-fold symmetry within the subregion, the quantity will be equal to 1. The results show that there is substantial nematic (n=2) local ordering with values of the order parameter ranging from 0.6 to 0.7 in a neighborhood of two particle lengths, indicating the tendency for neighboring rods to align along their long axes; the values for 3- and 4-fold symmetry are also appreciable. The quantity decays to approximately 0.1 when we average over the entire system. A fit to an exponential yields a 1/$e$ correlation length of approximately 8 particle lengths for nematic order -- nearly half the system size -- indicating medium-range order, intermediate between particle size and system size.

We have no firm explanation for the existence of propagating waves that are an order of magnitude faster than the particle speed.  One possibility is based on a Peclet number~\cite{Kessler}, defined as the Peclet number is the ratio of advection to diffusion $UL/D$ -- $U$, the particle speed; $L$, the particle length; and $D$, the diffusion constant found from the linear regime of the particle mean square displacements obtained by individually tracking the rods~\cite{Daniels2009}. For this argument, as an individual rod moves preferentially parallel to its long axis, the surrounding granular fluid becomes entrained and gives rise to advection flows. The entrained fluid causes neighboring rods to prefer to align in the direction of the advective flow and travel in the same direction. This is one way to understand the localized correlated motion observed in our system. When these localized domains collide with one another, the particles are constrained from moving but the entrained fluid flows past the particles, pushing the neighboring domain and resulting in a propagating wave. For our system, the Peclet number $UL/D$ -- using $U$ $\sim$ 1 cm/s and $D = 0.128$ cm$^{2}/$s  -- is approximately 10, suggesting that advection could play a role. A second possible explanation would entail an inertial argument in which the rods attempt to self-propel along their long axis but are constrained by the reduced free area available to them. Thus, during a collision with another rod, there is an additional transfer of momentum that pushes the second rod into a third rod, giving rise to a cascade of collisions that result in a propagating wave.

\section{Number Fluctuations}\label{nflucts}

\begin{figure}
\includegraphics[width=3.25in]{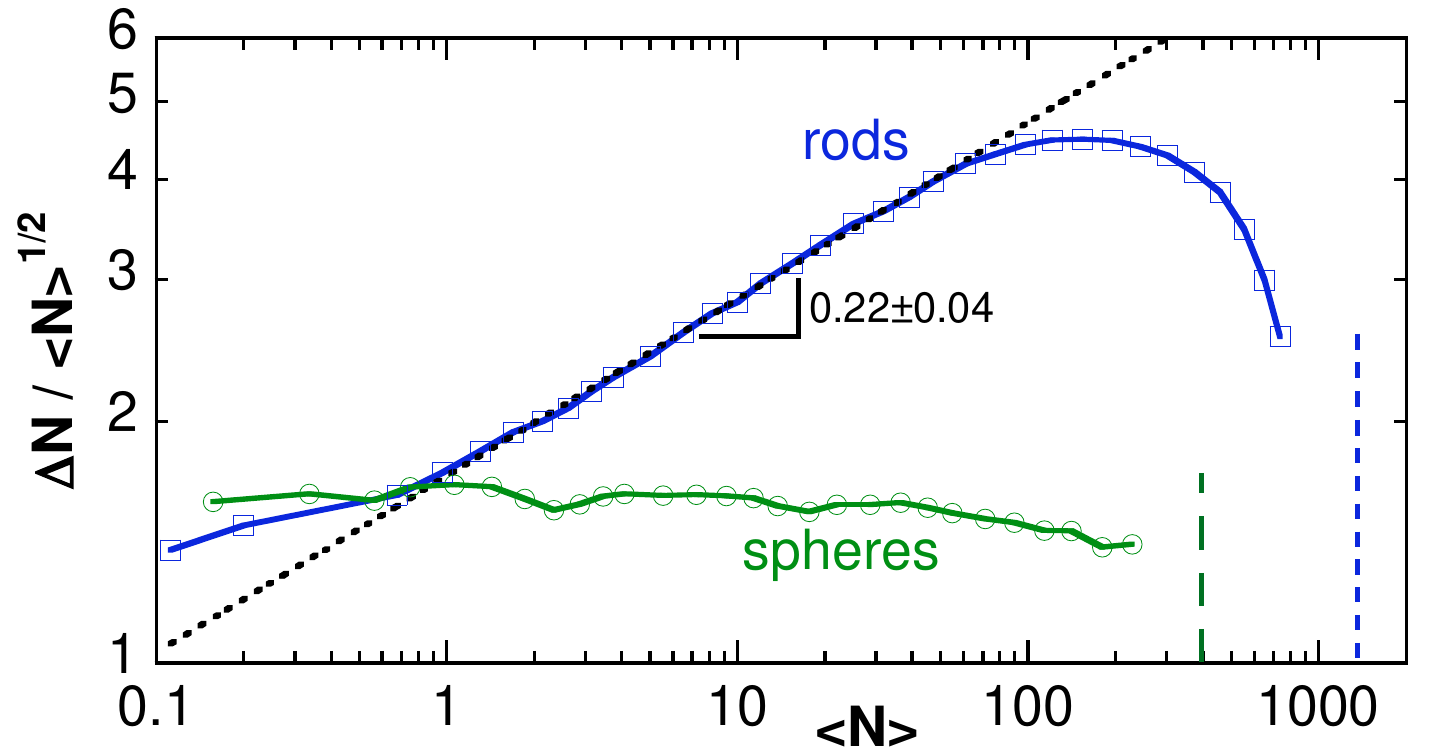}
\caption{(Color online) Magnitude of number fluctuations normalized by $\sqrt{\langle N\rangle}$ versus average number of particles in a subregion of the system. The circles (green) are for a bidisperse collection of spheres. The squares (blue) are for a collection of rods. Both systems occupy 64\% areal packing fraction and are fluidized at 220~cm/s. The vertical dashed lines correspond to the total number of particles in each system: 1353 rods and 400 spheres.  The dashed line is a power-law fit, which corresponds to $\Delta N \sim \langle N \rangle^{0.72\pm0.04}$.}\label{GNF}
\end{figure}

One interesting consequence of the propagating waves moving much faster than the individual rods is that compressed and dilute regions coexist for periods of time long compared to the time it takes the waves to propagate. Such persistent fluctuations in the density suggest the presence of anomalously large fluctuations in the local number density for these systems, which we now calculate. For a thermal system, the quantity $\Delta N / \sqrt{\langle N \rangle}$, where $\langle N \rangle$ is the mean number of particles in a subregion and $\Delta N$ is the standard deviation, should be a constant. Thus, thermal behavior will be characterized by a horizontal line on a plot of $\Delta N / \sqrt{\langle N\rangle}$ versus $\langle N \rangle$, whereas so-called giant number fluctuations will be characterized by a non-zero slope and a magnitude larger than one.

To quantify number fluctuations, we take a 10-minute video of the system at a given packing fraction and airflow. We select a square subregion of interest with side length $l$ and count the number of particles within that region for each frame. From this time series, we calculate $\langle N \rangle$ and $\Delta N$. We then repeat this procedure for systematically larger regions of interest, ranging from a single pixel up to roughly half of the system size.

Results for a system at $64\%$ packing fraction, fluidized at 220 cm/s, are shown in Figure~\ref{GNF}, plotting $\Delta N / \sqrt{\langle N\rangle}$ versus $\langle N \rangle$ for each subregion. The number fluctuations for bidisperse spheres show ordinary behavior, with $\Delta N/\sqrt{\langle N\rangle} \approx 1.5$ being constant over the entire range of subregion sizes. Rods, however, show number fluctuations with both an exponent and magnitude larger than thermal expectations, indicating giant fluctuations in local number density. The large $\langle N\rangle$ fall-off is a finite size effect; the vertical dashed lines in Fig.~\ref{GNF} indicate the total number of particles in each system. At its maximum prior to this fall-off, the fluctuations in local number density of the rods are larger than those for spheres by more than a factor of 3.  By fitting a power law to the rod data, we find that number fluctuations for the rods scale as $\Delta N \sim \langle N\rangle^{0.72\pm0.04}$.

Although the maximum in the number fluctuations is due to finite size effects, it serves as a useful benchmark to quantify the spatial extent of local number fluctuations. We accomplish this by converting the value of $\Delta N$ into an effective range of packing fractions.  The value of $N + \Delta N$ at the maximum corresponds to a range of packing fractions $\phi \sim 64\% \pm 20\%$. Thus, as the wave propagates through the system, particles are compressed by the front up to 84\% while the rarefaction zones are
diluted to approximately 44\%.

Figure~\ref{GNF} is strikingly similar to the number fluctuations obtained for rods fluidized by vertical vibrations~\cite{Menon2007}. In that experiment though, the large voids responsible for giant number fluctuations persisted for very long times, indicated by a logarithmic decay in the local density autocorrelation function. In our system, the density autocorrelation function decays to oscillations about zero within a few seconds.  These two systems exhibit qualitatively similar behavior -- giant number fluctuations -- caused by different mechanisms. The common factor between the two systems is that the rods self-propel at low densities.

\section{Waves and Density}\label{quantwaves}

Propagating waves have been predicted for collections of self-propelling particles interacting via volume exclusion with no emergent broken-symmetry state~\cite{Marchetti2008}, as well as more generally for particles adsorbed on a surface in a fluid background~\cite{Mazenko1982}.  In order to facilitate a theoretical description of our observations and understand the relative roles of particle shape and self-propelling, we must know how wavespeed and particle speed change with packing fraction and airflow. We do this by calculating a dynamic structure factor in subsection~\ref{DSF2}. We repeat the analysis for systems at different packing fractions and airflows in subsection~\ref{wspeed}.  Lastly, in subsection~\ref{phasebeh}, we map out the phase behavior of the rods.

\subsection{Density Power Spectrum}\label{DSF2}

\begin{figure}
\includegraphics[width=3.25in]{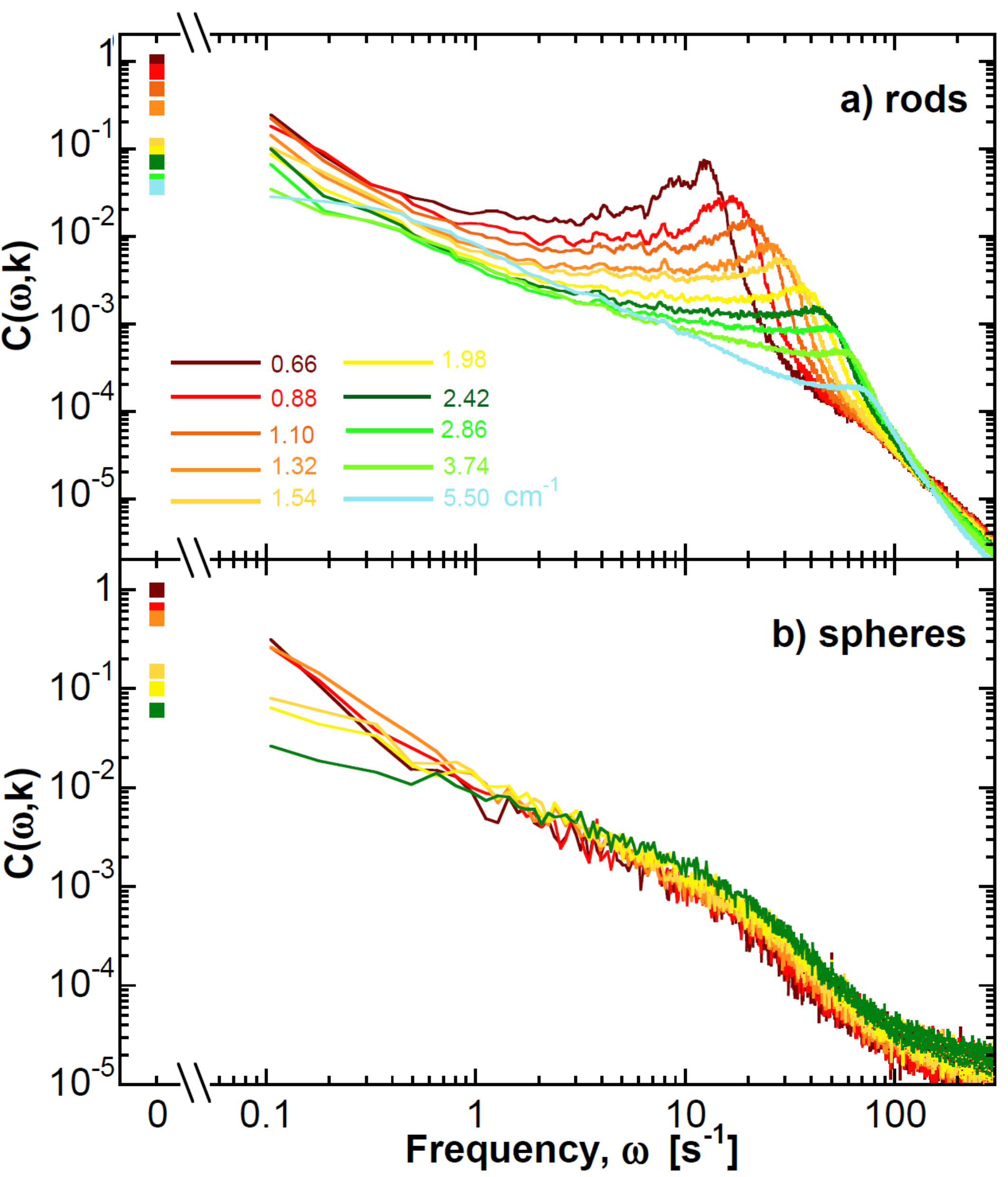}
\caption{(Color online) Spatiotemporal power spectrum of the density $C(\omega,k)$ for a) a collection of rods and b) a
collection of bidisperse spheres, both occupying 64\% packing fraction. Each curve corresponds to a different wave-vector value as shown by the color legend in a). The DC-component for each curve is plotted to the left of the axis break.}\label{DSF}
\end{figure}

In this subsection, we obtain more detailed quantitative information about the waves by calculating the spatiotemporal power spectrum $C(\omega,|k|)$ of the density $\rho (x,y,t)$ -- where $\omega$ is frequency and $|k|$ is wave-vector. Rather than compute the actual density, we simply use the 0-255 grayscale pixel values directly from the video data.  As seen in Fig.~\ref{WAVES}, this may be taken as a proxy for the local density. Using LabVIEW's Vision package, we first obtain the spatial Fourier transform of each frame in the video. Because there is no long-range order, the spatial Fourier transform of the rods data shows two rings at $|k| \simeq 4.96$~cm$^{-1}$ and 26.2~cm$^{-1}$, corresponding to the long and short dimensions of the rod. The annuli are isotropic with respect to the polar angle indicating that there is no long-range order in the system. We next extract the time-traces for all pixel values within an annulus at fixed $|k|$. We then  calculate the temporal power spectrum for each pixel time-trace at fixed $|k|$ and average over all pixels to obtain the spatiotemporal power spectrum $C(\omega,|k|)$. The magnitude of the power spectrum as both $\omega$ and $|k| \rightarrow 0$ is simply the sum of the grayscale values of each frame in the video averaged over time.  We normalize the power spectrum magnitude by this quantity so that the rods and spheres data can be more directly compared.

Slices of $C(\omega,|k|)$ for fixed $|k|$ are plotted versus $\omega$ in Fig.~\ref{DSF}(a) for rods and \ref{DSF}(b) for spheres, both at 64\% projected area fraction. The range of wave-vectors shown correspond to one-fourth the system size to approximately one particle length. The salient feature of $C(\omega,|k|)$ in Fig.~\ref{DSF}(a) is a single peak at a given $|k|$, indicating a traveling excitation. For larger wave-vectors outside of the hydrodynamic limit, the peak is no longer observed. In contrast, the same analysis for spheres in Fig.~\ref{DSF}(b) shows no such peaks.

\begin{figure}
\includegraphics[width=3.25in]{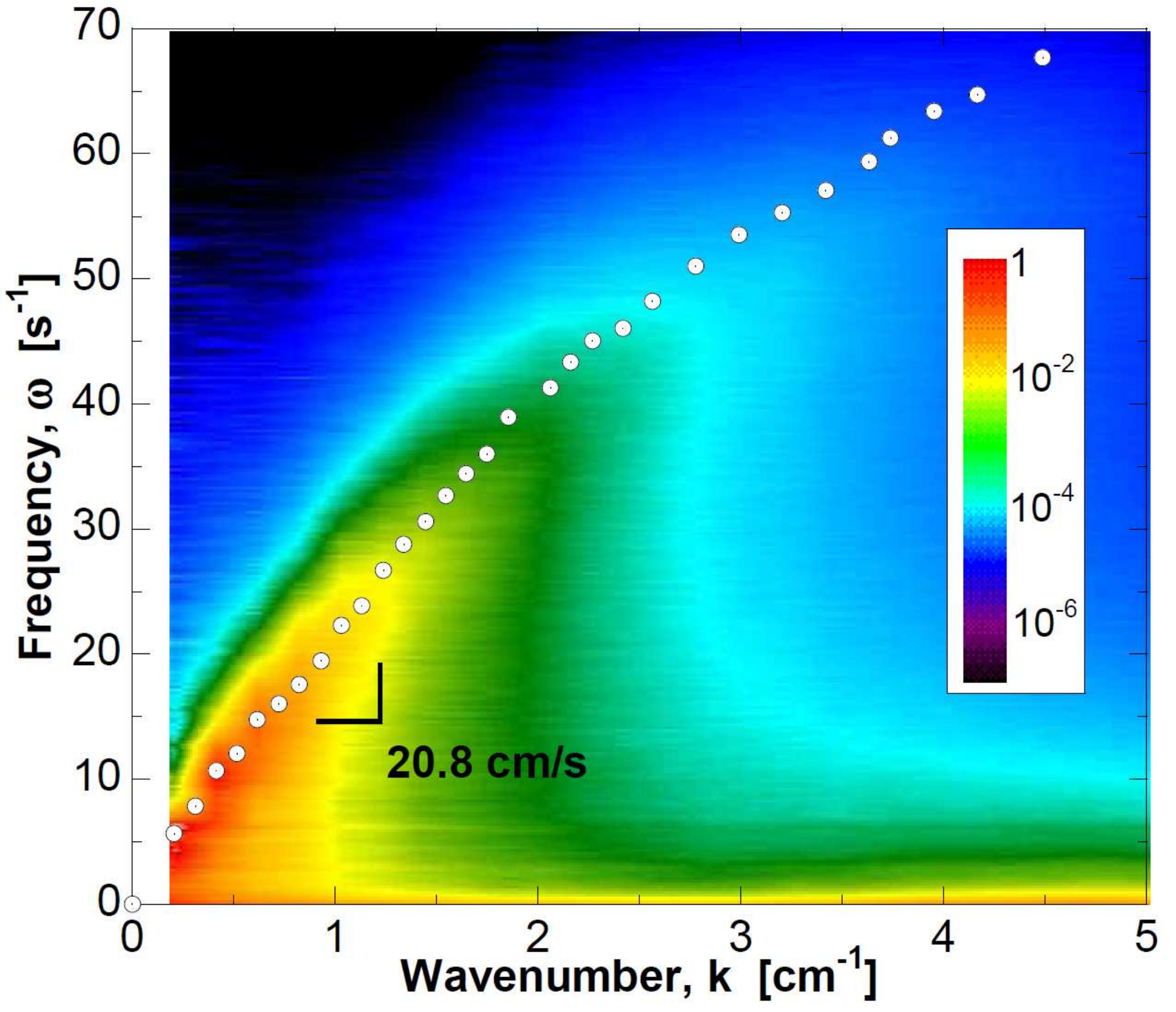}
\caption{(Color online) Contour plot of the spatiotemporal power spectrum of the density for rods
occupying 64\% packing fraction and fluidized at 220 cm/s. The power spectrum amplitude is given in the color legend. The
ocations of the peaks (circles) form the dispersion relationship
for the propagating waves. The slope of the linear region is the
wave speed $c =$ 20.8 cm/s.}\label{DISP}
\end{figure}

By plotting the location of peak frequency versus $|k|$, we can construct the dispersion relationship for the waves, as shown
by the symbols in Fig.~\ref{DISP}. On this same figure, we also superpose a colorized contour plot of $C(\omega,|k|)$. In the limit of small wavevector and frequency, the dispersion relation is linear and we may therefore extract a wave speed $c = \omega / |k|$ from the slope. For this particular example, $c=20.8$~cm/s.  This is consistent with the video data which show the waves propagating across the system in 1-2~seconds as well as the value obtained from spacetime plots. Again, we note that the wave speed is an order of magnitude larger than the
rms speed of the particles, $v_{rms} \sim 1$~cm/s, and an order of magnitude smaller than the fluidizing airflow $\sim200$~cm/s.

\subsection{Wavespeed}\label{wspeed}

\begin{figure}
\includegraphics[width=3.25in]{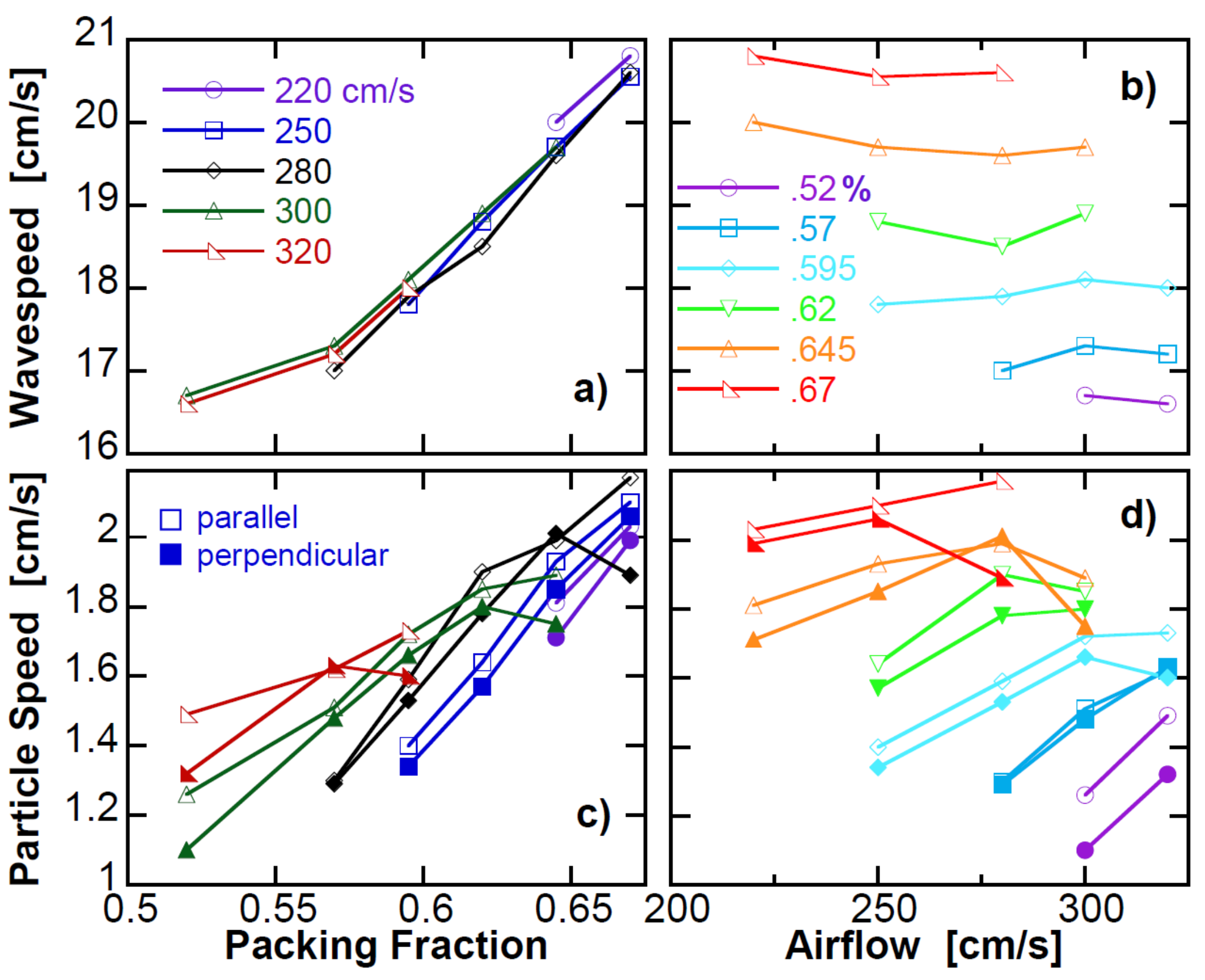}
\caption{(Color online) Wavespeed as a function of a) packing fraction and b) airflow and root-mean-square parallel (open symbols) and perpendicular (closed symbols) particle speeds as a function of c) packing fraction and d) airflow.}\label{ALLSPEED}
\end{figure}

\begin{figure}
\includegraphics[width=3.25in]{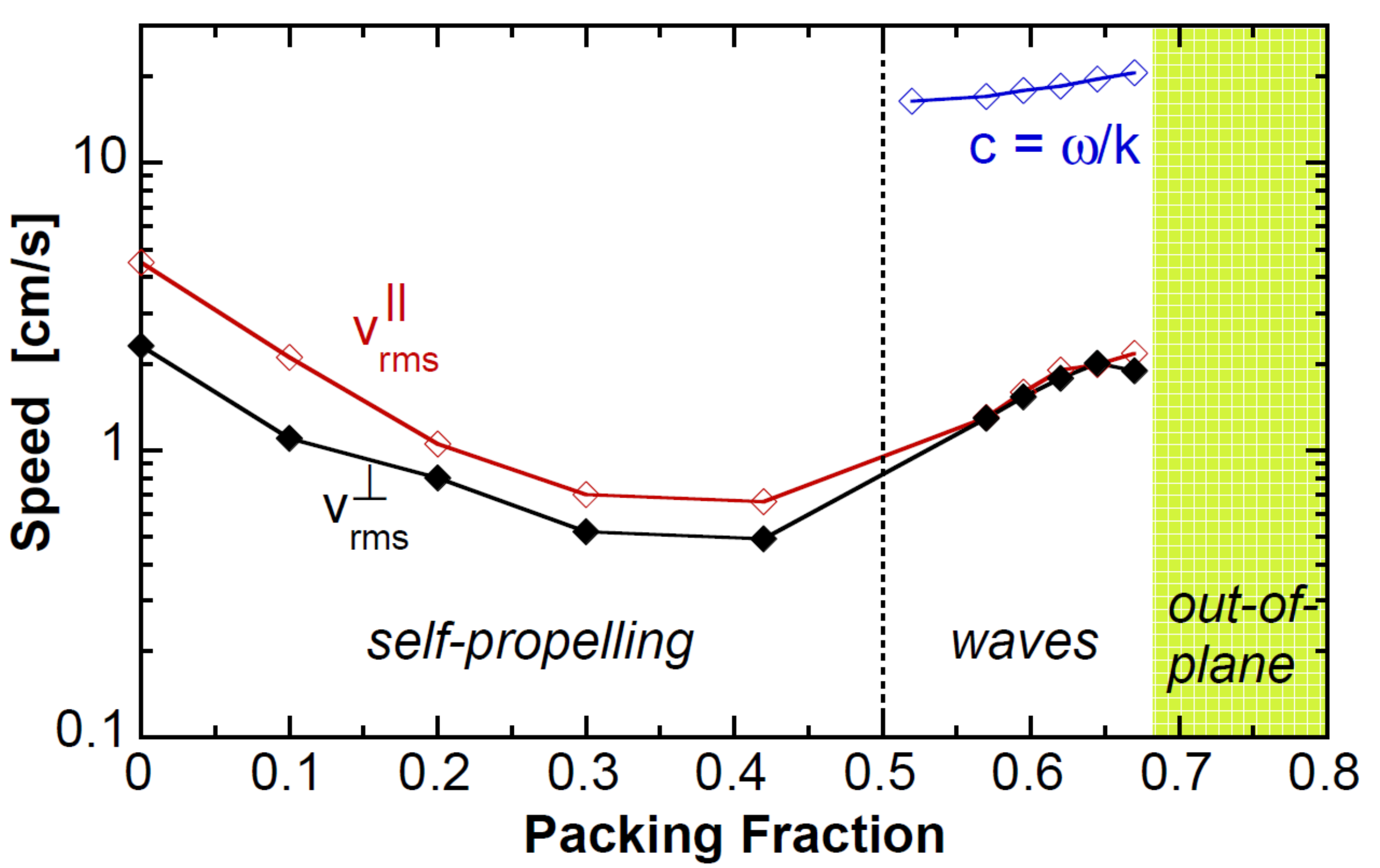}
\caption{(Color online) Wavespeed (blue circles), c = $\omega / k$, and root-mean-square parallel (red open diamonds) and perpendicular (black closed diamonds) particle speeds as a function of packing fraction for rods fluidized at 280 cm/s. The shaded green region is where out-of-plane motion becomes substantial.}\label{SPEEDS}
\end{figure}

Now we may measure the wavespeed as a function of both packing fraction and airspeed, using the spatiotemporal power spectra as illustrated in the examples of Figs.~\ref{DSF}-\ref{DISP}.  For comparison, we also use particle tracking to measure the rms speeds parallel and perpendicular to the long axis of the rods.  The results are displayed in the four plots of Fig.~\ref{SPEEDS}. At fixed air speed, the wave speed increases slightly as we increase the packing fraction of the system as seen in Fig.~\ref{ALLSPEED}(a). The increase appears to be roughly linear, although the dynamic range of our data is limited. This linear trend is seen for several other fixed airflows that we analyzed. By contrast, if we fix the packing fraction and increase the fluidizing airflow, the wavespeed shows no dependence on airflow, Fig.~\ref{ALLSPEED}(b). From Fig.~\ref{ALLSPEED}(c), we see that the particle speeds increase slightly with increasing density; from (d), we see a similar increase in particle speed with increasing airflow except at the highest airflows. For high airflows, out-of-plane motion dominates and in-plane motion slows down. For most densities, the parallel self-propelling speed is slightly larger than the transverse speed, although the effect is smaller than that observed for dilute systems~\cite{Daniels2009}. Although the rods are self-propelling, we see that the side-to-side motion induced by the propagating waves causes $v^{\perp}_{rms}$ to be nearly equal to $v^{\parallel}_{rms}$.

For all densities and airspeeds analyzed, the wavespeed remains an order of magnitude larger than the particle speed and the components of the particle speed remain nearly equal. This indicates that when propagating waves are present, the particles no longer self-propel. This is demonstrated more powerfully in Fig.~\ref{SPEEDS}, where we extend the particle rms speed measurements to lower packing fractions. For this particular density and airflow, waves are no longer observed below 50\% packing fraction. And below this packing fraction, we see that the parallel and perpendicular particle speeds become appreciably different. Thus, we conclude that self-propulsion and waves are exclusive of one another.

\subsection{Phase Behavior}\label{phasebeh}

\begin{figure}
\includegraphics[width=3.25in]{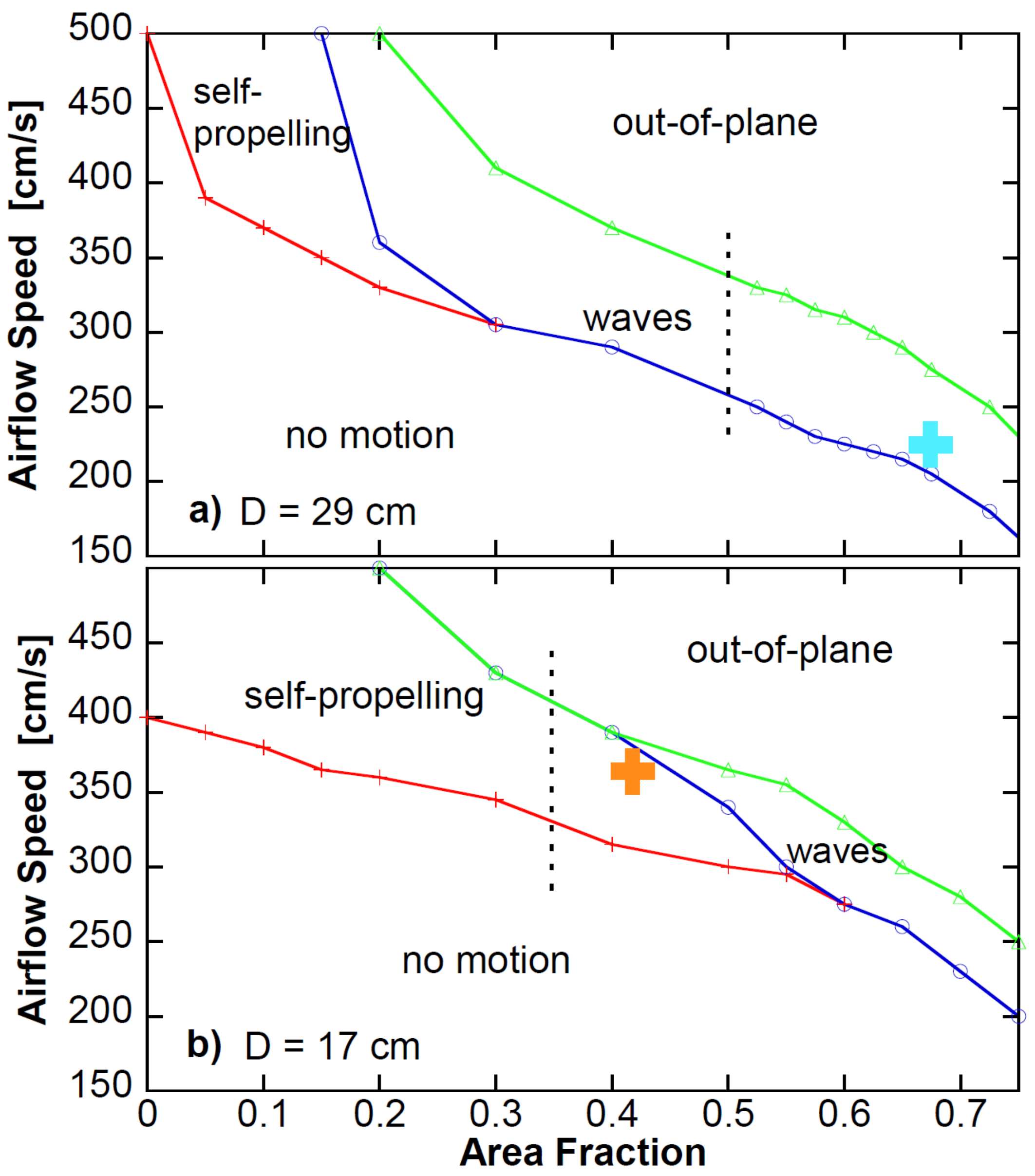}
\caption{(Color online) Behavior of fluidized rods as a function of fluidizing airflow and packing fraction. The diagram in a) is for a large system with diameter 29 cm; b) is for a smaller system with diameter 14 cm. The light blue cross in a) are the conditions for the analysis of Sections~\ref{quantwaves}A. The orange cross in b) represents the conditions for the experiment of Ref.~\cite{Daniels2009}. The vertical dashed line in both plots corresponds to the density above which the rods are uniformly distributed about the system.}\label{PHASE}
\end{figure}

Since we observed that self-propelling and waves are mutually exclusive, we want to determine the conditions for which a system of rods exhibits each phenomenon. Here, we qualitatively map out the phase behavior of fluidized rods as a function of packing fraction and fluidizing airflow, Fig.~\ref{PHASE}.  We explored a range of densities from a single particle up to 75\% packing fraction and fluidization speeds from 150 cm/s to 500 cm/s. We first place the desired number of rods into the system with no air flow. We then turn the airspeed to 150 cm/s. We increase the airspeed slowly by increments of roughly 10 cm/s. After each increase in the airflow, we wait 1 minute to ensure the system has reached a steady state and then characterize the behavior observed before increasing airflow once more. There are strong particle-wall interactions that cause the rods to cluster in the center of the system at low densities. Thus, the exact shape of the phase boundaries system-size-dependent. As such, we show a phase diagram for the full system size -- diameter 28.6 cm -- in Fig.~\ref{PHASE}(a) as well as the system size used in Ref.~\cite{Daniels2009} -- diameter 17 cm -- in Fig.~\ref{PHASE}(b). Because packing fraction is ill-defined unless the particles are uniformly distributed across the system, we denote the packing fraction above which the rods are uniformly dense as a vertical dashed line in both Figs.~\ref{PHASE}(a) and (b).

For all packing fractions, there is a threshold airflow below which the rods do not have enough energy to move. This is the lower boundary of the phase diagrams shown in Fig.~\ref{PHASE}.  Conversely, above some airflow, the rods gain enough energy to lose contact with the substrate and overlap one another out-of-plane. This marks the upper boundary of the phase diagrams which is determined by when we first observe two rods overlapping. Within these boundaries, we observe two distinct behaviors depending on both packing fraction and airflow. For dilute systems, the rods smoothly self-propel with very little out-of-plane motion. The experimental conditions for our previous work~\cite{Daniels2009} are shown as the solid orange cross in Fig.~\ref{PHASE}(b). As density is increased, there emerges an instability for compression waves to propagate. The onset of waves is chosen to be when ripples first appear to the eye.

\section{Conclusion}

We have investigated the behavior of dense collections of rods fluidized by an upflow of air. For a large range of densities and airflows, the monolayer of rods becomes unstable to the propagation of compression waves. The waves travel at a speed an order of magnitude larger than the root-mean-square speed of the particles themselves. As the compressed and rarefied regions relax after a wave's passage, the separation in timescales gives rise to anomalously large fluctuations in the local number density. Unlike giant number fluctuations expected for swarms and seen in vibrated-bed experiments, number fluctuations in our system are short-lived. Nevertheless, the particles in both experiments are self-propelling rods and these experiments suggest that self-propelling can be used as a useful concept to unify both physical and biological systems.

To ascertain the roles that self-propulsion and shape play in the collective wave behavior we observe, we quantified the waves by calculating the dynamic structure factor for the rods. From the position of peaks in this function, we extracted a dispersion relationship and, in the hydrodynamic limit, the wavespeed. Repeating our analysis for systems of increasing density and constant airflow, we found that the wavespeed is roughly linear with particle density and independent of airflow. Furthermore, for packing fractions at which waves are observed, the parallel and perpendicular rms particle speeds are the same, indicating that waves and self-propelling behavior are exclusive of one another for our system.

\vfil
\begin{acknowledgments}
We thank Aparna Baskaran for helpful discussions on particle and wave speeds, and we thank John Toner for helpful discussions on ordering and giant number fluctuations. This work was supported by the NSF through grant DMR-0704147.
\end{acknowledgments}

\bibliography{WaveRefs}

\end{document}